\documentclass[aps,pra,reprint,superscriptaddress,nofootinbib,floatfix]{revtex4-2}
\usepackage{graphicx}
\graphicspath{{Figure_pdf/}}
\usepackage{bm}
\usepackage{mathtools,amssymb}
\usepackage{physics}
\usepackage{xcolor}
\usepackage{placeins}
\usepackage{hyperref}
\usepackage{orcidlink}
\hypersetup{
    breaklinks,
    unicode,
    linktoc=all,
    bookmarksnumbered=true,
    bookmarksopen=true,
    colorlinks,
    linkcolor=blue,
    citecolor=blue,
    urlcolor=blue,
    plainpages=false,
    pdfstartview=FitH,
    pdfborder={0 0 0},
    linktocpage}

\makeatletter
\newcommand{\cofirstmark}{%
  \appdef\@affil@present@script{\textsuperscript{,\,\hyperlink{frontmatter.1}{*}}}%
}
\makeatother

\begin{document}

\title{Mitigating sloppiness with intermediate controls in sequential multiparameter spin interferometry}

\author{Hai Feng}
\thanks{These two authors contributed equally to this work.}
\affiliation{Beijing Academy of Quantum Information Sciences, Beijing 100193, China}
\affiliation{State Key Laboratory of Low Dimensional Quantum Physics, Department of Physics, Tsinghua University, Beijing 100084, China}

\author{Xinwei Li~\orcidlink{0009-0005-9677-0011}\texorpdfstring{\cofirstmark}{}}
\email[Contact author: ]{xinwei.li1991@outlook.com}
\affiliation{Beijing Academy of Quantum Information Sciences, Beijing 100193, China}

\author{Rong Lu}
\affiliation{State Key Laboratory of Low Dimensional Quantum Physics, Department of Physics, Tsinghua University, Beijing 100084, China}

\author{Li You~\orcidlink{0000-0002-3919-4768}}
\email[Contact author: ]{lyou@mail.tsinghua.edu.cn}
\affiliation{Beijing Academy of Quantum Information Sciences, Beijing 100193, China}
\affiliation{State Key Laboratory of Low Dimensional Quantum Physics, Department of Physics, Tsinghua University, Beijing 100084, China}
\affiliation{Frontier Science Center for Quantum Information, Beijing 100084, China}
\affiliation{Hefei National Laboratory, Hefei, Anhui 230088, China}

\date{\today}%

\begin{abstract}
Multiparameter quantum sensing requires different parameters to produce distinguishable probe responses while allowing their information to be extracted simultaneously. In sequential sensing without intermediate control, two phases encoded through the same generator are locally indistinguishable. Here we show that parameter-independent intermediate controls overcome this limitation in a collective-spin interferometer by separating the corresponding effective generators in the probe covariance geometry.
Specifically, a transverse collective rotation provides an analytically tractable realization, and intermediate one-axis twisting (OAT) generates additional nonlinear operator directions and a broad regime of enhanced joint estimation.
For fixed total particle number $N$ and matched twisting resource, both marginal precision bounds lie below those of a separated-OAT strategy while measurement incompatibility remains weak. For intermediate OAT, conditional optimization of an upper bound on the Holevo cost yields asymptotic $N^{-2}$ scaling. For this protocol, a fixed rotated-Dicke measurement retains simultaneous two-parameter enhancement over a broad phase region. These results show that nonlinear intermediate control can mitigate sloppiness while preserving quantum-enhanced precision with a common readout.
\end{abstract}

\maketitle

\section{Introduction}\label{sec:intro}
Quantum metrology uses nonclassical probe states to improve measurement precision~\cite{Giovannetti_2006,Giovannetti_2011,Degen_2017,Pezze_2018}. In collective-spin systems, one-axis twisting (OAT) generates spin-squeezed and non-Gaussian states that enable phase estimation beyond the standard quantum limit (SQL)~\cite{Wineland_1992,Kitagawa_1993,Ma_2011,Strobel2014Fisher,Pezze_2018}, with metrologically useful correlations demonstrated in Bose gases and atomic ensembles~\cite{Gross_2010,Riedel_2010,Muessel_2015,Colombo_2022}. Collective rotations and nonlinear interactions also provide control over probe preparation and interaction-based readout~\cite{Davis_2016,Schulte_2020,Carrasco_2022,Scharnagl_2023}. A central question for multiparameter sensing is how these controls can be used when several unknown quantities are encoded in the same quantum probe.

Many sensing tasks involve multiple unknown quantities, such as phases, vector-field components, or time-dependent signals~\cite{Tsang_2011,Humphreys_2013,Yue_2014,Crowley_2014,Zhuang2020,Kaubruegger_2023}. Sloppiness reflects a strong disparity in sensitivity to different parameter combinations and is characterized by an ill-conditioned quantum Fisher information matrix (QFIM)~\cite{Waterfall_2006,Gutenkunst_2007,Transtrum_2015}. In the singular limit, some combinations become locally unidentifiable~\cite{Goldberg_2021,Mihailescu_2026}. Measurement incompatibility presents a distinct limitation: a nonsingular QFIM does not guarantee simultaneous attainability of the symmetric-logarithmic-derivative (SLD) Cram\'{e}r--Rao bounds for different parameters~\cite{Helstrom_1976,Holevo_1982,Matsumoto_2002,Ragy_2016,Szczykulska_2016,Albarelli_2019,Albarelli_2020,Candeloro_2024}. A simple fixed readout need not approach the quantum precision bounds, even when incompatibility is weak.

In sequential sensing, different parameters may be encoded through the same probe--signal coupling during successive time intervals. When two phase shifts are generated by the same operator without an intermediate control, the final state depends only on their sum, making the two phases individually unidentifiable. Quantum control can remove this degeneracy by modifying the evolution between the encoding stages~\cite{Yang2022Untwining}. Related approaches include control and scrambling protocols in discrete- and continuous-variable systems~\cite{Frigerio_2024,He_2025,Sharma_2025,Manju_2026}, as well as intermediate or repeated measurements~\cite{Bizzarri_2025,Yang_2025_Sequential}. These studies establish intermediate intervention as a means of restoring parameter distinguishability. The question addressed here is how intermediate control in a collective-spin interferometer can restore identifiability while retaining quantum-enhanced precision and allowing both phases to be estimated with a common fixed readout.

We answer this question using an OAT-prepared ensemble of $N$ spin-$1/2$ particles that acquires two phases through successive $J_z$ rotations. We consider two distinct protocols that differ in the parameter-independent control applied between the two encodings. For both protocols, the QFIM is proportional to the real covariance matrix of the local effective generators evaluated on the prepared probe. Local identifiability requires these generators to produce independent local responses on that probe; noncommutation of the intermediate control with $J_z$ alone is therefore insufficient. In the first protocol, we use a transverse collective rotation as an analytically tractable reference: it reorients the second effective generator within the linear spin operator space. In the second protocol, an intermediate OAT operation introduces nonlinear spin terms into this generator, providing a distinct mechanism for separating the two phase responses.

We compare both protocols with two independent $N/2$-particle OAT probes under matched particle and twisting resources. In suitable control regimes, both marginal SLD bounds lie below this separated-OAT reference, while weak incompatibility keeps the Holevo bound close to the SLD cost. For intermediate OAT, conditional optimization of an upper bound on the Holevo cost gives finite-size behavior consistent with $N^{-2}$ scaling. Finally, we evaluate the classical Fisher information matrix (CFIM) for fixed rotated-Dicke measurements~\cite{Dicke_1954,Lucke_2011,Zhang_2014,Zou_2018}. With these readouts, intermediate OAT lowers the classical marginal bounds for both phases over a broad sampled phase region, whereas the linear protocol provides simultaneous improvement over a more limited region. This comparison shows that favorable QFIM geometry alone does not ensure simultaneous enhancement over a broad phase region; such enhancement also depends on how effectively the chosen readout extracts information about both phases.

Section~\ref{sec:def} introduces the estimation framework and the measures of local identifiability, precision, and measurement incompatibility. Section~\ref{sec:mitigating_slopp} presents the sequential sensing scheme, defines the separated benchmarks, and analyzes the linear and nonlinear intermediate controls. Section~\ref{sec:dicke_readout} evaluates the precision attainable with fixed Dicke-basis readouts. Section~\ref{sec:conclusion} summarizes the main results and discusses possible extensions.

\section{Multiparameter quantum estimation framework}\label{sec:def}
We consider the joint estimation of the two phases
$\bm{\theta}=(\theta_1,\theta_2)^T$ encoded in a quantum state
$\rho_{\bm{\theta}}$. At the quantum-state level, the local information
about $\bm{\theta}$ is characterized by the QFIM. The SLD operator $L_i$
associated with $\theta_i$ is defined implicitly by
\begin{equation}
    \partial_i\rho_{\bm{\theta}}
    =
    \frac{1}{2}
    \left(
        \rho_{\bm{\theta}}L_i
        +
        L_i\rho_{\bm{\theta}}
    \right),
    \label{eq:SLD_def}
\end{equation}
where $\partial_i\equiv\partial/\partial\theta_i$. The QFIM is then
given by~\cite{Helstrom_1976,Holevo_1982,Albarelli_2020}
\begin{equation}
    Q_{ij}
    =
    \frac{1}{2}
    \operatorname{Tr}
    \left[
        \rho_{\bm{\theta}}
        \{L_i,L_j\}
    \right],
    \label{eq:QFIM_SLD}
\end{equation}
where $\{A,B\}=AB+BA$. For nonsingular $Q$, any locally unbiased
estimator $\hat{\bm{\theta}}$ and any real positive-definite weight
matrix $W$ satisfy the scalar SLD bound
\begin{equation}
    \operatorname{Tr}
    \left[
        W\Sigma(\hat{\bm{\theta}})
    \right]
    \geq
    C_Q(W)
    \equiv
    \operatorname{Tr}
    \left(
        WQ^{-1}
    \right),
    \label{eq:CQ_def}
\end{equation}
where $\Sigma(\hat{\bm{\theta}})$ is the covariance matrix, with the
common inverse repetition factor suppressed in the bounds.
We assign equal weight to the two phases and set $W=\mathbb I_2$.
Accordingly,
$C_Q=\operatorname{Tr}(Q^{-1})$, while $(Q^{-1})_{ii}$ gives the SLD
marginal bound for $\theta_i$. Unlike in single-parameter estimation,
these SLD bounds need not be simultaneously attainable.

The rank of the QFIM determines whether all parameter directions are
locally identifiable. Because $Q$ is positive semidefinite, a singular
$Q$ has a nonzero null direction $\bm{v}$ satisfying
$\bm{v}^{T}Q\bm{v}=0$. An infinitesimal displacement along this
direction therefore produces no locally distinguishable change in the
quantum state. If $Q$ is nonsingular but strongly anisotropic, a small
eigenvalue instead identifies a poorly resolved parameter direction.
For the two-parameter models considered here, we characterize the
approach to rank deficiency using the determinant-based sloppiness
diagnostic~\cite{Candeloro_2024}
\begin{equation}
    \mathcal{S}
    =
    \frac{1}{\det Q}.
    \label{eq:sloppiness_def}
\end{equation}
Within the parameterization adopted below, $\mathcal{S}$ diverges as
$\det Q$ approaches zero, signaling the loss of local identifiability.

A nonsingular QFIM ensures local identifiability but does not guarantee
that the SLD bounds for different parameters can be attained
simultaneously. This limitation arises from measurement
incompatibility and is characterized by the Uhlmann-curvature matrix
\begin{equation}
    U_{ij}
    =
    -\frac{i}{2}
    \operatorname{Tr}
    \left[
        \rho_{\bm{\theta}}[L_i,L_j]
    \right],
    \label{eq:U_def}
\end{equation}
together with the normalized incompatibility measure~\cite{Carollo_2019,Razavian_2020,Candeloro_2024}
\begin{equation}
    \mathcal R
    =
    \left\|
        iQ^{-1}U
    \right\|_{\infty},
    \qquad
    0\leq\mathcal R\leq1,
    \label{eq:R_def}
\end{equation}
where $\|\cdot\|_{\infty}$ denotes the largest eigenvalue modulus.
For nonsingular $Q$, $\mathcal R=0$ is equivalent to the
weak-commutativity condition $U=0$. The Holevo Cram\'er--Rao bound
$C_H(W)$ gives the optimal asymptotic weighted precision cost when
collective measurements across independent copies of
$\rho_{\bm{\theta}}$ are allowed~\cite{Albarelli_2019,DemkowiczDobrzanski_2020}. For pure-state models, the Holevo bound can be attained asymptotically using suitable measurements on individual copies~\cite{Matsumoto_2002,Albarelli_2020}.
$C_H(W)$ accounts for measurement incompatibility and satisfies
\begin{equation}
    C_Q(W)
    \leq
    C_H(W)
    \leq
    C_Q(W)(1+\mathcal R).
    \label{eq:Holevo_bracket}
\end{equation}
Thus, a small $\mathcal R$ tightly brackets the Holevo bound near the
SLD cost without requiring its explicit evaluation. For a nonsingular
two-parameter model, the antisymmetry of $U$ gives
$\det U=U_{12}^{\,2}$ and hence
\begin{equation}
    \mathcal R
    =
    \sqrt{\frac{\det U}{\det Q}}
    =
    \sqrt{\frac{\mathcal{S}}{\mathcal{C}}},
    \qquad
    \mathcal{C}
    =
    \frac{1}{\det U}.
    \label{eq:R_two_parameter}
\end{equation}
The inverse-curvature diagnostic $\mathcal C$ characterizes the compatibility condition and it diverges when the
Uhlmann curvature vanishes, whereas $\mathcal S$ diverges as the QFIM
approaches rank deficiency. The incompatibility measure $\mathcal R$
is defined only for nonsingular $Q$.

These quantities distinguish local identifiability, SLD precision, and
measurement incompatibility. For the unitary pure-state protocols below,
we evaluate them using the covariances of the local effective generators.
Section~\ref{sec:mitigating_slopp} uses this representation to analyze
how a transverse collective rotation and intermediate OAT modify the two
phase responses, and how these changes affect the precision bounds and
measurement incompatibility.

\begin{figure}
    \centering
    \includegraphics[width=1.0\linewidth]{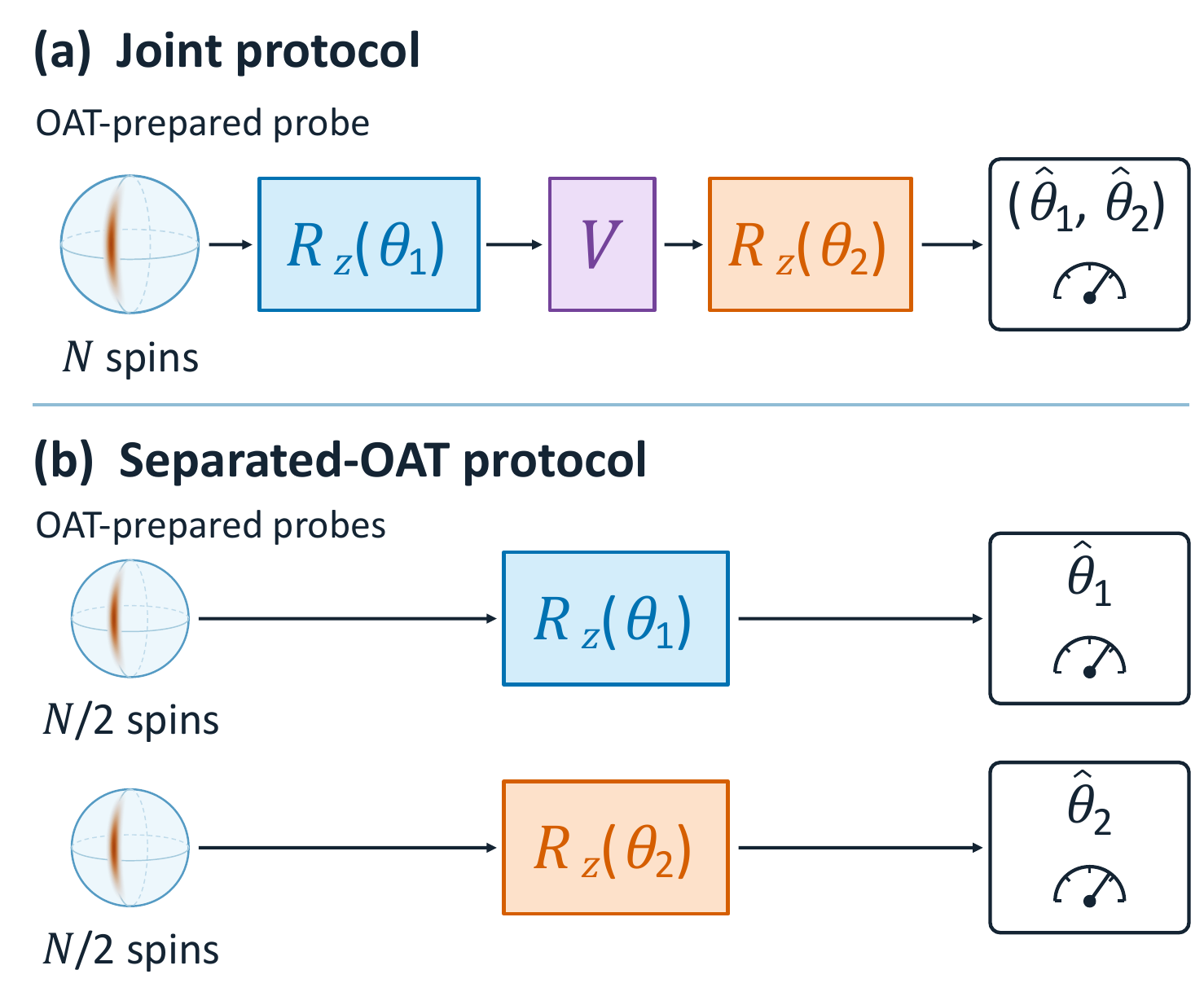}
    \caption{Joint and separated-OAT estimation protocols. (a) An OAT-prepared $N$-particle probe encodes $\theta_1$ and $\theta_2$ through $R_z(\theta_1)$ and $R_z(\theta_2)$, with a parameter-independent intermediate control $V$, chosen as a transverse collective rotation or an OAT operation. (b) Two independent OAT-prepared $N/2$-particle probes estimate the phases separately. Each separated probe is twisted for the total OAT duration of the corresponding joint protocol.
    }
    \label{fig:protocol}
\end{figure}

\section{Mitigation of sloppiness with intermediate controls}\label{sec:mitigating_slopp}

\subsection{Sensing scheme and benchmarks}
We consider an ensemble of $N$ spin-$1/2$ particles restricted to the
permutation-symmetric subspace with total spin $J=N/2$. The collective
spin operators are
\begin{equation}
    J_\mu
    =
    \frac{1}{2}
    \sum_{n=1}^{N}
    \sigma_\mu^{(n)},
    \qquad
    \mu=x,y,z,
\end{equation}
where $\sigma_\mu^{(n)}$ is the Pauli operator of particle $n$ along
direction $\mu$. For a unit vector $\bm n$, we define
$J_{\bm n}=\bm n\cdot\bm J$ and denote collective rotations and OAT
operations by
\begin{equation}
    R_{\bm n}(\varphi)
    =
    e^{-i\varphi J_{\bm n}},
    \qquad
    T_{\bm n}(t)
    =
    e^{-i\chi tJ_{\bm n}^{2}},
\end{equation}
where $\chi$ is the nonlinear interaction strength and $t$ is the
interaction duration. The unknown phases $\theta_1$ and $\theta_2$
are dimensionless.

The initial coherent spin state is polarized along $+x$,
$\ket{\psi_0}=\ket{J,m_x=J}$. We prepare the probe by applying OAT
about the $z$ axis for a duration $t_1$, followed by a rotation about
the $x$ axis:
\begin{equation}
    \ket{\psi_P}
    =
    R_x(\delta)T_z(t_1)\ket{\psi_0}.
\end{equation}
The rotation $R_x(\delta)$ aligns the anti-squeezed quadrature with the
$z$ axis, thereby maximizing the variance relevant to the subsequent
$J_z$-generated phase encoding. The required alignment angle is
\begin{equation}
    \delta
    =
    \frac{\pi}{2}
    -
    \frac{1}{2}
    \arctan\!\left(\frac{B}{A}\right),
\end{equation}
where $A = 1-\cos^{N-2}(2\chi t_1)$ and $B = 4\sin(\chi t_1)\cos^{N-2}(\chi t_1)$~\cite{Kitagawa_1993}.

The two phases are encoded sequentially through rotations generated by
$J_z$. After the first encoding $R_z(\theta_1)$, we apply a parameter-independent control $V$, followed by the second encoding
$R_z(\theta_2)$. The final state is
\begin{equation}
    \ket{\psi_{\bm{\theta}}}
    =
    R_z(\theta_2)
    V
    R_z(\theta_1)
    \ket{\psi_P}.
    \label{eq:final_state_scheme}
\end{equation}
As illustrated in Fig.~\ref{fig:protocol}, the joint protocol uses one
$N$-particle probe to estimate both phases. We consider two choices of
the intermediate control: a transverse collective rotation and an OAT
operation of duration $t_2$. These controls are analyzed in
Secs.~\ref{subsec:linear} and~\ref{subsec:nonlinear}, respectively.

To calculate the QFIM and the Uhlmann-curvature matrix for this
protocol, we introduce the local generators of the two phase
parameters. Writing
$\mathcal U_{\bm{\theta}}=R_z(\theta_2)VR_z(\theta_1)$,
we define the effective Hermitian local generator associated with $\theta_i$, with respect to the prepared probe $\ket{\psi_P}$, by
\begin{equation}
\mathcal H_i
=
i\mathcal U_{\bm{\theta}}^\dagger
\partial_i\mathcal U_{\bm{\theta}}.
\end{equation}
These generators express both infinitesimal phase responses in a common
frame referred back to the prepared probe. The first encoding acts
directly on this probe, while the second is preceded by
$V R_z(\theta_1)$, so its generator is conjugated by these operations:
\begin{equation}
\mathcal H_1
=
J_z,
\qquad
\mathcal H_2
=
R_z^\dagger(\theta_1)
V^\dagger J_z V
R_z(\theta_1).
\label{eq:local_generators_SS}
\end{equation}
We can therefore evaluate both matrices using covariances in
$\ket{\psi_P}$. With
$\langle X\rangle_P=\mel{\psi_P}{X}{\psi_P}$ and
$\operatorname{Cov}_P(X,Y)
=\langle XY\rangle_P-\langle X\rangle_P\langle Y\rangle_P$,
the QFIM and Uhlmann curvature of the encoded pure state are~\cite{Liu_2020}
\begin{subequations}
\label{eq:Q_U_generator_SS}
\begin{align}
Q_{ij}
&=
4\,\operatorname{Re}
\operatorname{Cov}_P(\mathcal H_i,\mathcal H_j),
\\
U_{ij}
&=
4\,\operatorname{Im}
\operatorname{Cov}_P(\mathcal H_i,\mathcal H_j).
\end{align}
\end{subequations}
The real covariance matrix determines sensitivity and local
identifiability, while the imaginary part captures the average
noncommutativity relevant to measurement incompatibility.

For $V=\mathbb I$, the two local generators coincide,
$\mathcal H_1=\mathcal H_2=J_z$, and the final state depends only on
the accumulated phase $\theta_1+\theta_2$. The QFIM then reduces to
\begin{equation}
    Q
    =
    4\operatorname{Var}_P(J_z)
    \begin{pmatrix}
        1 & 1\\
        1 & 1
    \end{pmatrix},
\end{equation}
which has rank one and cannot distinguish the two phase contributions.
Changing the second local direction requires $[V,J_z]\neq0$, but
noncommutation alone does not guarantee rank lifting. For the
two-parameter model, the QFIM is nonsingular if
\begin{equation}
    \operatorname{Var}_P(\mathcal H_1)
    \operatorname{Var}_P(\mathcal H_2)
    -
    \left[
        \operatorname{Re}
        \operatorname{Cov}_P(\mathcal H_1,\mathcal H_2)
    \right]^2
    >0.
\end{equation}
Thus, both generators must have nonzero variance on the prepared
probe, and their fluctuations must not be perfectly correlated.

We compare all protocols at the same total particle number $N$. In the
separated SQL, two independent $N/2$-particle coherent probes estimate
the phases individually, giving marginal and total variance bounds
$2/N$ and $4/N$, respectively. For the separated-OAT reference, we
also match the particle-weighted twisting strength
$N\chi t_{\mathrm{tot}}$, where $t_{\mathrm{tot}}=t_1$ for the
linear-rotation protocol and $t_{\mathrm{tot}}=t_1+t_2$ for the
intermediate-OAT protocol. Each reference probe is twisted for
$t_{\mathrm{tot}}$, so the two branches together use the same twisting
resource as the joint probe. Collective rotations are treated as ideal
controls and are not included in this twisting resource.

For each separated-OAT branch, the preparation and alignment follow
the same construction with $N$ replaced by $N/2$ and $t_1$ by
$t_{\mathrm{tot}}$. We denote its single-parameter QFI by
$q_{\mathrm{sep}}=4\operatorname{Var}_{\mathrm{sep}}(J_z)$, where both
the variance and $J_z$ refer to that $N/2$-particle probe.
The separated-OAT marginal and total variance bounds are then
$1/q_{\mathrm{sep}}$ and $2/q_{\mathrm{sep}}$.
The total-cost gains and the joint marginal gains relative to the
separated SQL are
\begin{subequations}
\label{eq:quantum_gains}
\begin{align}
G_{\mathrm{sep}}
&=10\log_{10}\left(\frac{2q_{\mathrm{sep}}}{N}\right),
\\
G_{\mathrm{joint}}
&=10\log_{10}\left(\frac{4}{NC_Q}\right),
\\
G_i
&=10\log_{10}\left[\frac{2}{N(Q^{-1})_{ii}}\right],
\qquad i=1,2.
\label{eq:marginal_gain}
\end{align}
\end{subequations}
All gains are reported in decibels. For the two identical separated-OAT
probes, $G_{\mathrm{sep}}$ is also the marginal gain for either phase.
A positive gain indicates a variance bound below the separated SQL.
The joint total cost improves on the resource-matched
separated-OAT strategy when $G_{\mathrm{joint}}>G_{\mathrm{sep}}$;
the corresponding marginal condition is $G_i>G_{\mathrm{sep}}$.

\subsection{Linear-rotation protocol}\label{subsec:linear}
\begin{figure}
    \centering
    \includegraphics[width=1.0\linewidth]
    {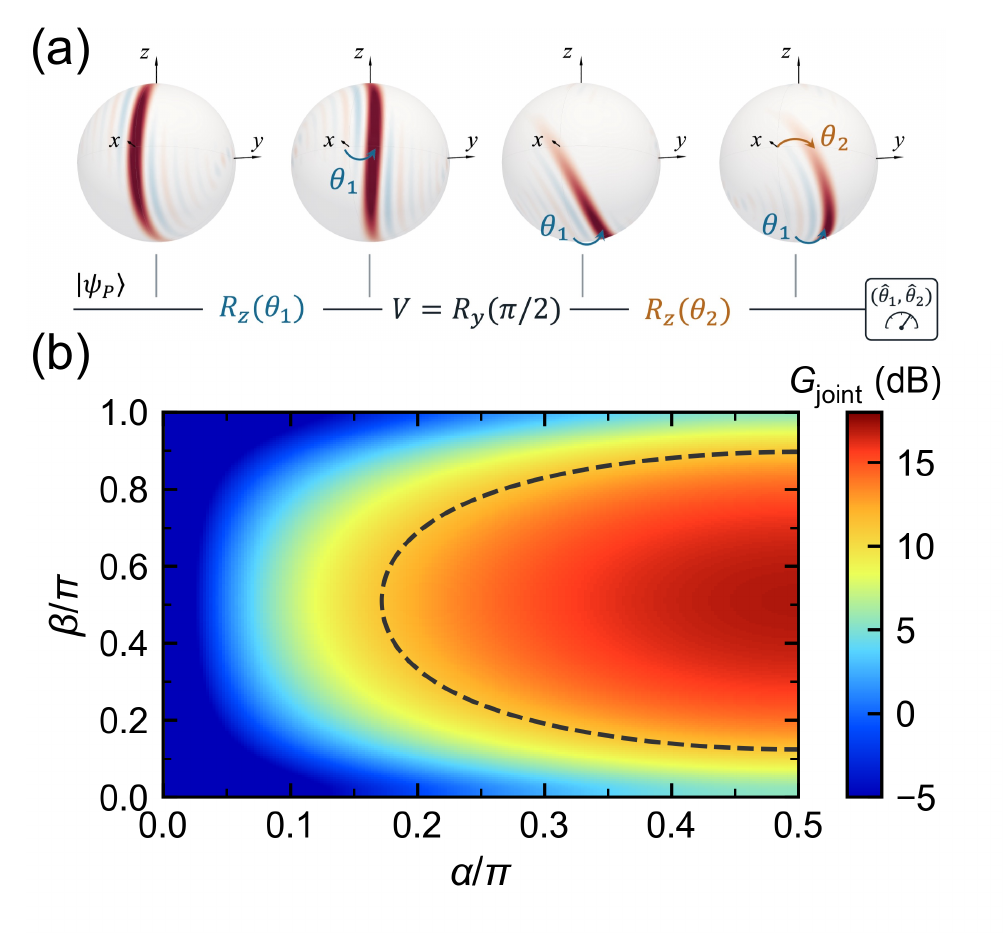}
    \caption{
    Linear-rotation protocol and its metrological gain.
    (a) Illustrative spin Wigner distributions along the encoding
    sequence with intermediate control $R_y(\pi/2)$.
    The encoding angles in (a) are exaggerated for visibility.
    (b) Joint gain $G_{\mathrm{joint}}$ relative to the separated SQL
    versus $\alpha$ and $\beta$.
    The dashed contour marks equality with the resource-matched
    separated-OAT reference $G_{\mathrm{sep}}$.
    The QFIM is singular at $\alpha=0$.
    In (b), $N=50$, $\chi t_1=0.1\pi$, and $\theta_1=0.01\pi$.
    }
    \label{fig:CQ_heatmap_rotation}
\end{figure}

We first consider an intermediate collective rotation about an axis in
the $xy$ plane,
\begin{equation}
V
=
R_{\beta}(\alpha)
=
e^{-i\alpha J_{\beta}},
\qquad
J_{\beta}
=
J_x\cos\beta+J_y\sin\beta,
\label{eq:V2_linear}
\end{equation}
where $\alpha$ is the rotation angle and $\beta$ specifies the
orientation of the rotation axis. Substituting this control into
Eq.~\eqref{eq:local_generators_SS} gives
\begin{equation}
\begin{split}
\mathcal H_2
&=J_z\cos\alpha-J_x\sin\alpha\sin(\beta-\theta_1)
\\
&\quad+J_y\sin\alpha\cos(\beta-\theta_1).
\end{split}
\label{eq:linear_generators}
\end{equation}
The angle $\alpha$ controls the transverse component of $\mathcal H_2$,
while $\beta$ selects the spin variance sampled by that component.
The angle $\beta-\theta_1$ is measured relative to the probe after
the first phase encoding. At $\alpha=\pi/2$, the second generator is
transverse to $J_z$, and the two parameter directions are uncorrelated
in the real covariance matrix of the aligned probe.

For the prepared probe, the relevant first moments and real cross
covariances satisfy $\langle J_y\rangle_P=\langle J_z\rangle_P=0$ and $\operatorname{Re}\operatorname{Cov}_P(J_x,J_y)=\operatorname{Re}\operatorname{Cov}_P(J_x,J_z)=\operatorname{Re}\operatorname{Cov}_P(J_y,J_z)=0$. Writing $\mathcal V_\mu=\operatorname{Var}_P(J_\mu)$, with $\mu=x,y,z$, the independent QFIM elements at $\theta_1=0$ are
\begin{subequations}
\label{eq:linear_Q_elements}
\begin{align}
Q_{11}
&=
4\mathcal V_z,
\\
Q_{12}
&=
Q_{21}
=
4\mathcal V_z\cos\alpha,
\\
Q_{22}
&=
4\left[
\mathcal V_z\cos^2\alpha
+
\sin^2\alpha
\left(
\mathcal V_x\sin^2\beta
+
\mathcal V_y\cos^2\beta
\right)
\right].
\end{align}
\end{subequations}
The first diagonal element is unchanged by the control, whereas the
off-diagonal element is proportional to $\cos\alpha$ and vanishes at
$\alpha=\pi/2$. Suppressing this off-diagonal coupling can improve the
marginal bound for $\theta_1$, because that bound is determined by the
inverse of the full QFIM. The transverse part of the second generator
contributes the additional variance in $Q_{22}$ and supplies the
information needed to distinguish $\theta_2$ from $\theta_1$. The
Uhlmann-curvature matrix is antisymmetric, with
$U_{11}=U_{22}=0$ and $U_{21}=-U_{12}$. Its only independent element
at $\theta_1=0$ is
\begin{equation}
U_{12}
=
-2\langle J_x\rangle_P\sin\alpha\cos\beta .
\label{eq:linear_U_element}
\end{equation}
The corresponding expressions at arbitrary $\theta_1$ are given in
Appendix~\ref{app:linear_rotation}.

At $\theta_1=0$, the determinant obtained from
Eq.~\eqref{eq:linear_Q_elements} is
\begin{equation}
\det Q
=
16\mathcal V_z\sin^2\alpha
\left(
\mathcal V_x\sin^2\beta
+
\mathcal V_y\cos^2\beta
\right).
\label{eq:linear_detQ}
\end{equation}
For $\mathcal V_z>0$, the determinant is positive when
$\sin\alpha\ne0$ and the selected transverse spin component has nonzero
variance on the probe. Restoring full rank establishes local
identifiability. The precision bounds also depend on the magnitude of
the probe fluctuations along the transverse direction. In the
preparation regime $\mathcal V_x>\mathcal V_y$, the SLD cost at
$\theta_1=0$ is minimized by making the second generator transverse to
$J_z$ and directing it along $J_x$:
\begin{equation}
\alpha
=
\frac{\pi}{2}+k\pi,
\qquad
\beta
=
\frac{\pi}{2}+\ell\pi,
\qquad
k,\ell\in\mathbb Z.
\label{eq:linear_optimal_angles}
\end{equation}
The complete minimization for arbitrary control angles is given in
Appendix~\ref{app:linear_rotation}.

\begin{figure}
    \centering
    \includegraphics[width=1.0\linewidth]
    {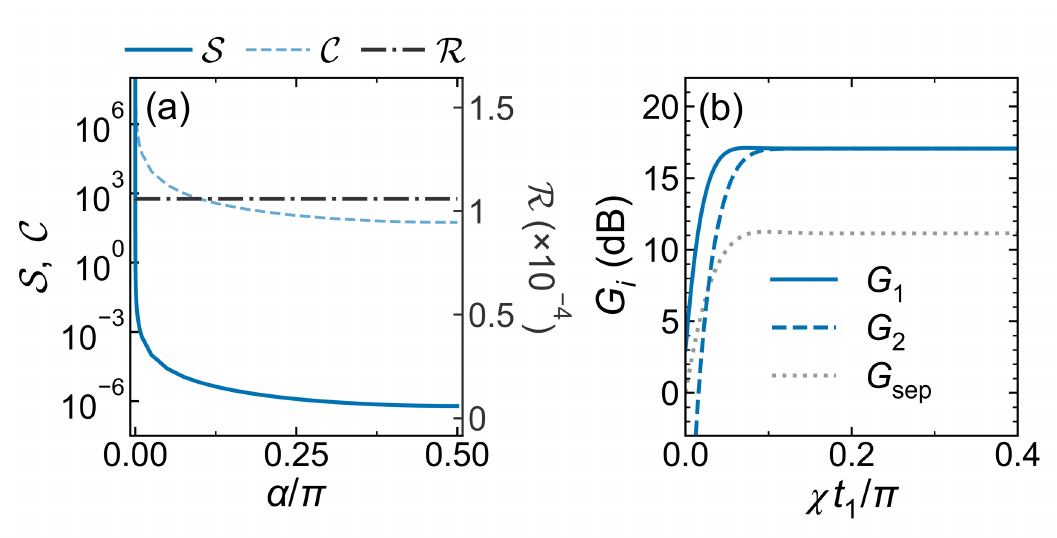}
    \caption{
    Performance of the linear-rotation protocol.
    (a) Sloppiness diagnostic $\mathcal S$ and inverse-curvature
    diagnostic $\mathcal C$ (left axis), and incompatibility measure
    $\mathcal R$ (black dash-dotted curve; right axis, in units of
    $10^{-4}$), versus $\alpha/\pi$, with $\chi t_1=0.1\pi$.
    The QFIM is singular at $\alpha=0$.
    (b) Marginal gains $G_1$ and $G_2$ relative to the separated SQL
    versus $\chi t_1/\pi$, with $\alpha=\pi/2$.
    The gray dotted curve denotes the separated-OAT QFI reference
    $G_{\mathrm{sep}}$.
    Here $\beta=\pi/2$, $N=50$, and $\theta_1=0.01\pi$.
    }
    \label{fig:Qinv_rotation}
\end{figure}

Figure~\ref{fig:CQ_heatmap_rotation} summarizes the operation and
performance of the linear-rotation protocol. Panel~(a) shows that the
intermediate rotation reorients the probe between the two phase
encodings, allowing $\theta_1$ and $\theta_2$ to produce
distinguishable state variations. Panel~(b) shows the corresponding
joint gain over the $(\alpha,\beta)$ plane. At $\alpha=0$, the control
is the identity, independent of $\beta$, and the QFIM is singular.
The gain increases as the second generator is rotated away from $J_z$
and is maximized near $\alpha=\beta=\pi/2$, where the transverse
component probes the larger variance $\mathcal V_x$. The dashed
contour marks $G_{\mathrm{joint}}=G_{\mathrm{sep}}$, showing that the
joint protocol outperforms the resource-matched separated-OAT
strategy over a finite range of control angles.

Figure~\ref{fig:Qinv_rotation}(a) shows how the information geometry
changes with $\alpha$. For $\beta=\pi/2$, the curvature vanishes at
$\theta_1=0$; the small nonzero phase used in
Fig.~\ref{fig:Qinv_rotation}(a) makes $\mathcal C$ finite away from
$\alpha=0$. As $\alpha\rightarrow0$, the effective generators become
identical, making the QFIM singular and causing $\mathcal S$ to
diverge. The Uhlmann curvature also vanishes in this limit, causing
$\mathcal C$ to diverge. Increasing $\alpha$ rotates the second
generator away from $J_z$, thereby separating the two parameter
directions and decreasing $\mathcal S$. At the same time,
$\mathcal R$ remains small throughout the nonsingular region shown,
indicating that the Holevo cost stays close to the SLD cost. The
curvature magnitude and $\sqrt{\det Q}$ have the same $|\sin\alpha|$
dependence, explaining the constant $\mathcal R$ along this scan.

Figure~\ref{fig:Qinv_rotation}(b) complements the total-cost comparison
in Fig.~\ref{fig:CQ_heatmap_rotation} by resolving the gains for the
two phases separately. It shows the marginal gains defined in
Eq.~\eqref{eq:marginal_gain} at $\alpha=\beta=\pi/2$. At these control
angles, the QFIM is diagonal, so the two marginal gains are determined
directly by the corresponding diagonal elements. The first is
governed by $Q_{11}=4\mathcal V_z$, while the second is governed by
$Q_{22}\simeq4\mathcal V_x$ near $\theta_1=0$. Because
$\mathcal V_x$ is initially small, $G_2$ lies below $G_1$ at short
twisting times. As OAT increases $\mathcal V_x$, the two gains approach
the same plateau and both exceed the separated-OAT reference
indicated by the gray dotted curve~\cite{Kitagawa_1993,Pezze_2018}.
Moreover, $Q_{22}=4\mathcal V_x+
4(\mathcal V_y-\mathcal V_x)\theta_1^2+O(\theta_1^4)$, while
$Q_{11}$ is independent of $\theta_1$. Thus, the first marginal gain
is independent of $\theta_1$, while the leading correction to the
second is quadratic near $\theta_1=0$.

\subsection{Intermediate-OAT protocol}\label{subsec:nonlinear}
\begin{figure}
    \centering
    \includegraphics[width=1.0\linewidth]{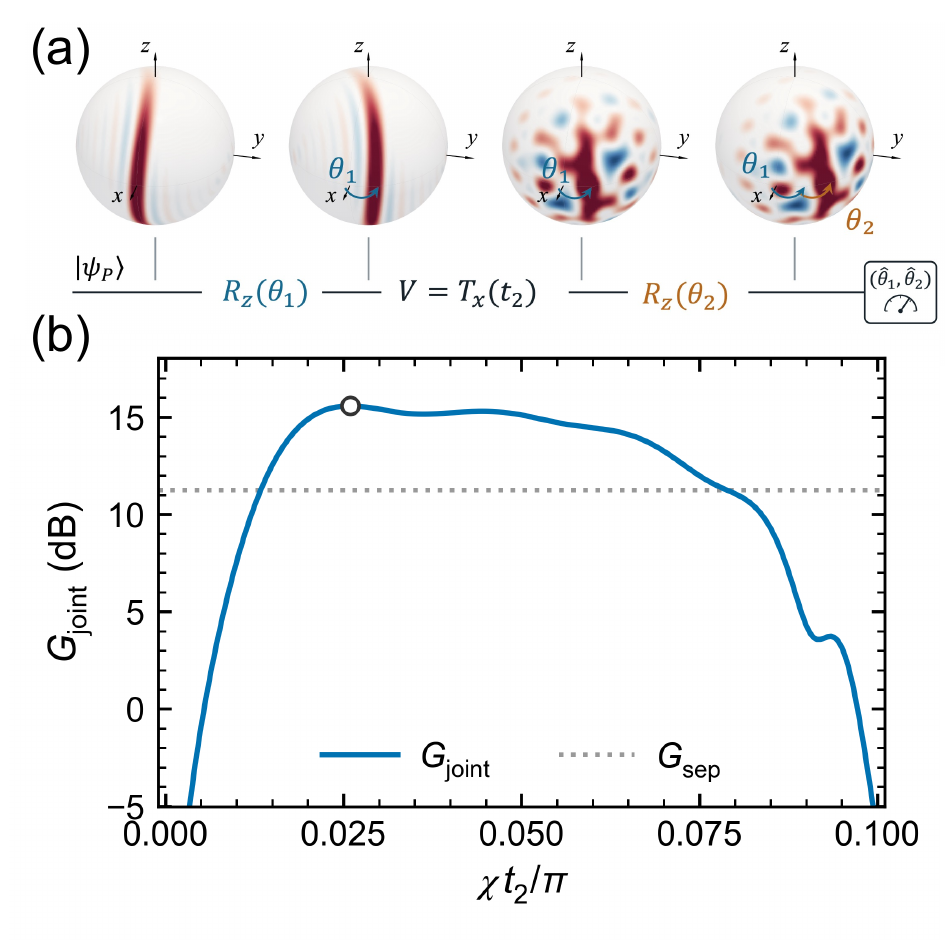}
    \caption{
    Intermediate-OAT protocol and joint precision gain.
    (a) Spin Wigner distributions of the prepared probe and its evolution
    under the successive operations $R_z(\theta_1)$, $T_x(t_2)$, and
    $R_z(\theta_2)$, from left to right. The encoding angles are exaggerated
    for visibility.
    (b) Joint gain $G_{\mathrm{joint}}$ relative to the separated SQL versus
    $\chi t_2/\pi$, at fixed $\chi(t_1+t_2)=0.1\pi$.
    The gray horizontal dotted line denotes the resource-matched
    separated-OAT gain $G_{\mathrm{sep}}$. The open circle marks the maximum
    over the sampled values of $t_2$, whose preparation and control settings
    are used in Fig.~\ref{fig:CFI}. In (b), $N=50$ and $\theta_1=0.01\pi$.
    }    
\label{fig:CQ_heatmap}
\end{figure}

We next consider an intermediate OAT operation about the $x$ axis,
\begin{equation}
V
=
T_x(t_2)
=
e^{-i\chi t_2J_x^2}.
\label{eq:V2_twisting}
\end{equation}
The linear-rotation protocol keeps $\mathcal H_2$ linear in the spin
components, whereas intermediate OAT introduces products of spin
operators. We use twisting about $x$ as a representative choice with a
simple leading correction; Appendix~\ref{app:twisting} examines other
axes in the $xy$ plane. At $\theta_1=0$, a short-time expansion gives
\begin{equation}
\mathcal H_2
=J_z+\chi t_2\{J_x,J_y\}+O[(\chi t_2)^2].
\label{eq:short_time_generator}
\end{equation}
The expansion applies for
$|\chi t_2|J\ll1$. The leading correction is first order in $\chi t_2$
and quadratic in the spin operators, and is absent from the first phase
generator $\mathcal H_1=J_z$. When the fluctuations of $\{J_x,J_y\}$
and $J_z$ are linearly independent in the real covariance metric of the
prepared probe, this correction makes $\det Q$ positive at order
$(\chi t_2)^2$, restoring local distinguishability.
Appendix~\ref{app:weak_oat} gives the full expansion and the corresponding
determinant. The numerical calculations below use the exact local
generators.

\begin{figure}
    \centering
    \includegraphics[width=1.0\linewidth]{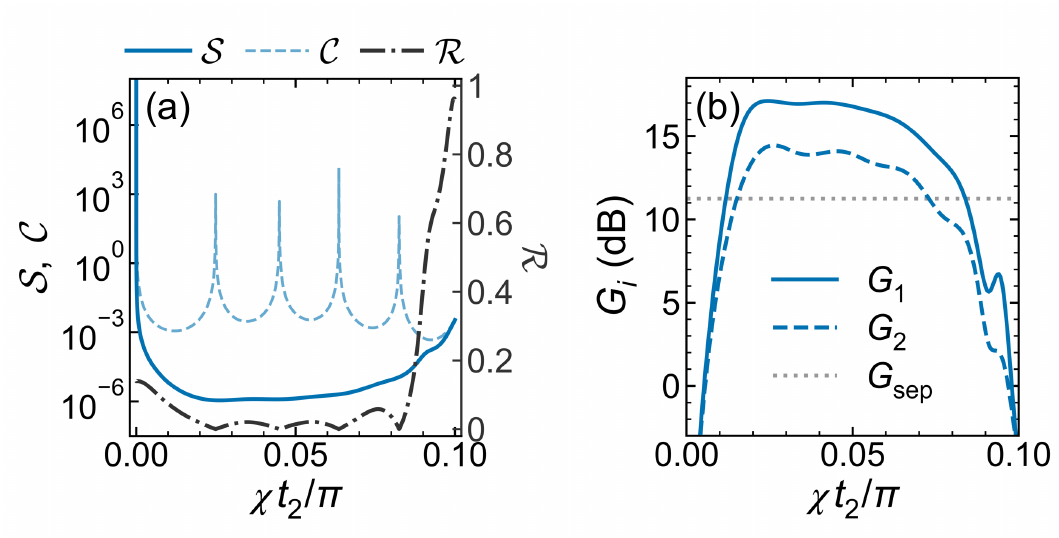}
    \caption{
    Information geometry and marginal gains of the intermediate-OAT protocol.
    (a) Sloppiness diagnostic $\mathcal S=1/\det Q$ and inverse-curvature
    diagnostic $\mathcal C=1/\det U$ on the logarithmic left axis, and
    incompatibility measure $\mathcal R$ on the linear right axis, versus
    $\chi t_2/\pi$. The black dash-dotted curve represents $\mathcal R$.
    At the interior divergences of $\mathcal C$, the Uhlmann curvature
    vanishes while the QFIM remains nonsingular.
    (b) Marginal gains $G_1$ for $\theta_1$ (solid) and $G_2$ for $\theta_2$
    (dashed), relative to the separated SQL. The gray horizontal dotted
    line denotes the resource-matched separated-OAT gain $G_{\mathrm{sep}}$.
    The scan keeps $\chi(t_1+t_2)=0.1\pi$ fixed.
    }
    \label{fig:Qinv}
\end{figure}

\begin{figure}
    \centering
    \includegraphics[width=1.0\linewidth]{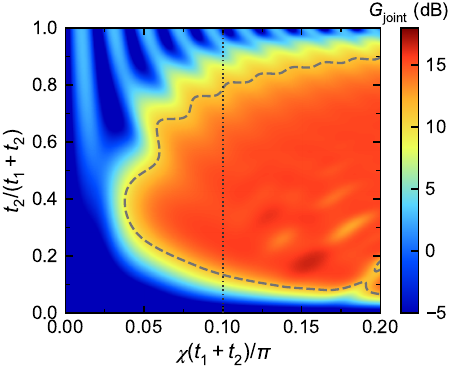}
    \caption{
    Joint gain $G_{\mathrm{joint}}$ of the intermediate-OAT protocol,
    relative to the separated SQL, versus the total twisting strength
    $\chi(t_1+t_2)/\pi$ and allocation ratio $t_2/(t_1+t_2)$, for $N=50$.
    The gray dashed contour marks $G_{\mathrm{joint}}=G_{\mathrm{sep}}$.
    At each total twisting strength, the separated-OAT reference is
    evaluated with the same total particle number and twisting resource.
    The black vertical dotted line at $\chi(t_1+t_2)/\pi=0.1$ marks the
    fixed-total-time slice shown in Figs.~\ref{fig:CQ_heatmap}(b)
    and~\ref{fig:Qinv}.
    }

\label{fig:squeezing_time_heatmap}
\end{figure}

Figure~\ref{fig:CQ_heatmap}(a) illustrates the nonlinear deformation
of the probe during intermediate OAT. At fixed total twisting time $t_{\mathrm{tot}}=t_1+t_2$,
Fig.~\ref{fig:CQ_heatmap}(b) shows a broad interval of joint gain above
the resource-matched separated-OAT reference. The reference remains
fixed along this scan, so the variation in joint gain reflects the
allocation of the same twisting resource. For small $t_2$, the two
phase responses remain poorly distinguished. As $t_2$ approaches
$t_{\mathrm{tot}}$, the remaining preparation time
$t_1=t_{\mathrm{tot}}-t_2$ becomes insufficient to generate enhanced
probe fluctuations. The maximum therefore balances the fluctuations
generated during preparation with the separation of the two responses
produced by intermediate control. 

Figure~\ref{fig:Qinv} resolves the corresponding information geometry
and marginal gains. At $t_2=0$, the QFIM is singular and the Uhlmann
curvature vanishes, so both $\mathcal S$ and $\mathcal C$ diverge.
The interior divergences of $\mathcal C$ instead arise from additional
curvature zeros. The QFIM remains nonsingular at these points, where
$\mathcal R=0$. Over most of the interval of enhanced sensitivity,
$\mathcal S$ is small and $\mathcal R$ remains close to zero, so
$C_Q$ and $C_Q(1+\mathcal R)$ tightly bracket the Holevo cost.
Both marginal gains exceed the separated-OAT reference over an
extended interval, with the gain for $\theta_1$ generally larger.
As $t_2$ increases further, the shorter preparation time reduces both
gains. The symmetry of the probe and control under a $\pi$ rotation
about $x$ makes the QFIM even in $\theta_1$, so its leading correction
near $\theta_1=0$ is quadratic.

Figure~\ref{fig:squeezing_time_heatmap} extends this comparison to
different total twisting strengths and allocations between the two
stages. The separated-OAT reference is evaluated at the corresponding
total twisting strength. A broad region exceeds this matched reference
when both stages have sufficient duration. The edge $t_2=0$ has a
singular QFIM, whereas $t_1=0$ corresponds to an initially coherent
probe without OAT preparation. The vertical dotted line gives the
fixed-total-time slice used in Figs.~\ref{fig:CQ_heatmap}
and~\ref{fig:Qinv}.

\begin{figure}
    \centering
    \includegraphics[width=1.0\linewidth]{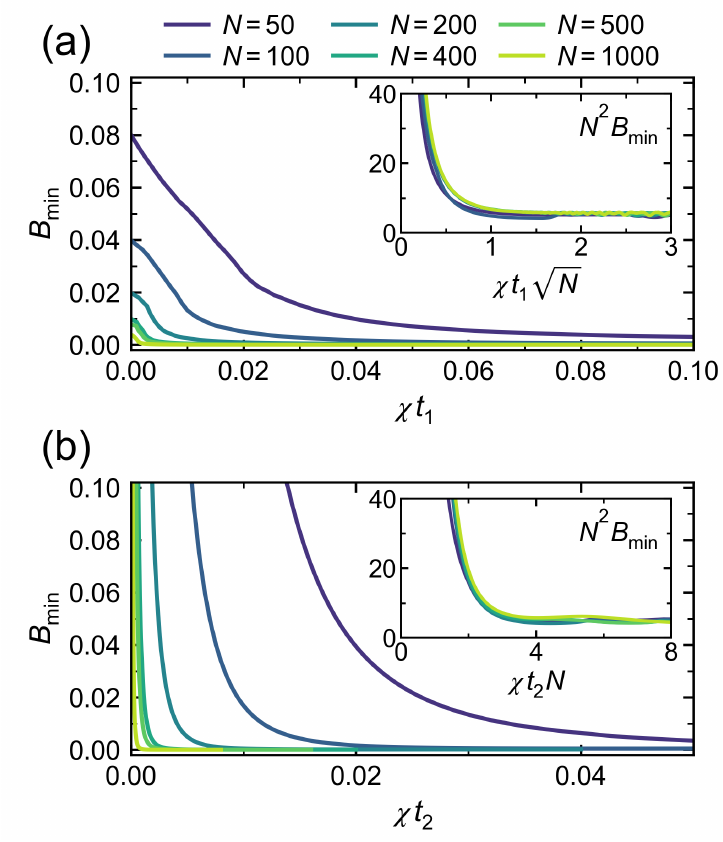}
    \caption{
    Particle-number scaling of the intermediate-OAT protocol.
    $B_{\min}$ denotes the conditional minimum of the Holevo upper bound
    $C_Q(1+\mathcal R)$, evaluated over the sampled interval of the
    optimized twisting time.
    (a) $B_{\min}$ versus $\chi t_1$, with $t_2$ optimized at each fixed $t_1$.
    (b) $B_{\min}$ versus $\chi t_2$, with $t_1$ optimized at each fixed $t_2$.
    The total twisting time varies during these optimizations. Different
    colors correspond to the particle numbers listed in the legend.
    Insets: $N^2B_{\min}$ versus $\chi t_1\sqrt N$ in (a) and $\chi t_2N$
    in (b). The approximately overlapping plateaus demonstrate
    $B_{\min}\propto N^{-2}$. Here $\theta_1=0.01\pi$.
    }
    \label{fig:Scaling}
\end{figure}

We next investigate the particle-number scaling by optimizing the
upper bound $C_Q(1+\mathcal R)$ on the Holevo cost. For each $N$,
we hold one twisting time fixed, minimize this bound over the sampled
interval of the other, and denote the result by $B_{\min}$.
In Fig.~\ref{fig:Scaling}(a), $t_2$ is optimized at each fixed $t_1$;
in Fig.~\ref{fig:Scaling}(b), $t_1$ is optimized at each fixed $t_2$.
The total twisting time varies during these optimizations, allowing
the dependence on the preparation and intermediate-control durations
to be examined separately.

Figure~\ref{fig:Scaling} numerically demonstrates an $N^{-2}$ scaling
of the optimized Holevo upper bound in the plateau regime. In the
insets, the rescaled curves approximately collapse, with $N^2B_{\min}$
approaching a common plateau. The horizontal rescalings identify
characteristic preparation and intermediate-control scales
$\chi t_1\sim N^{-1/2}$ and $\chi t_2\sim N^{-1}$, respectively.
These scalings show that the plateau is reached at shorter durations
as $N$ increases, with the characteristic intermediate-control time
decreasing more rapidly than the preparation time.

\begin{figure*}
    \centering
    \includegraphics[width=1.0\linewidth]{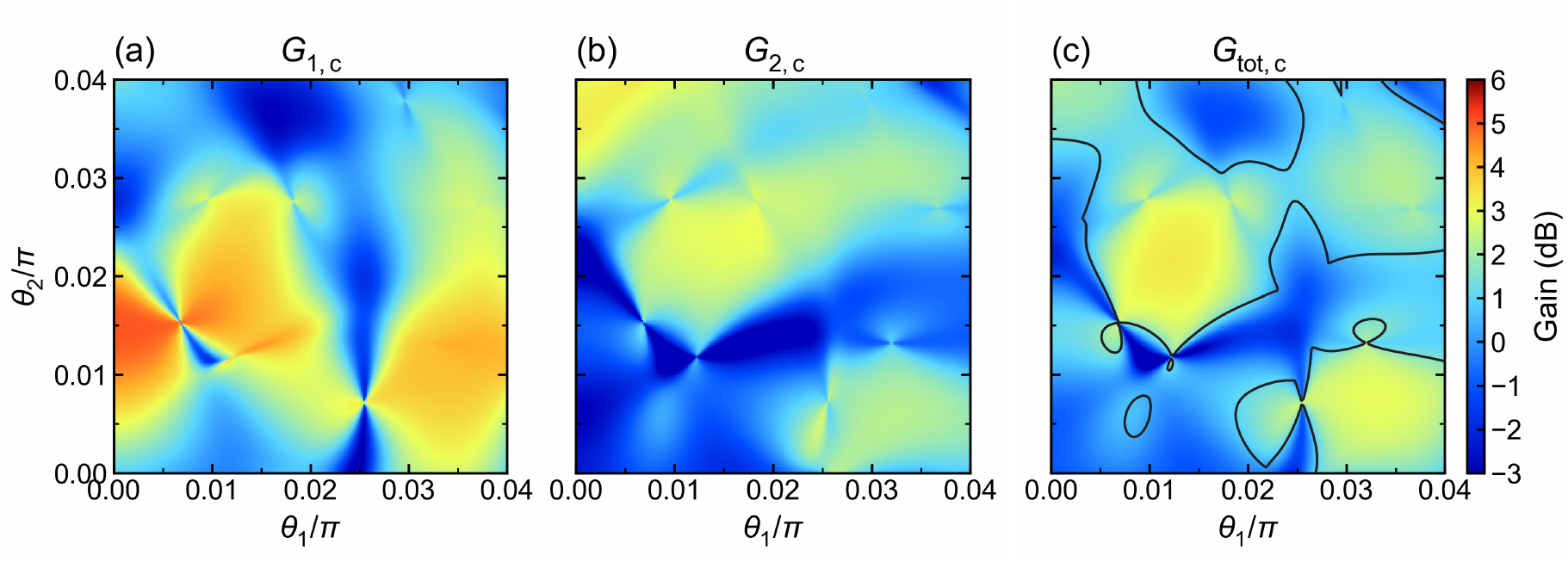}
    \caption{
    Classical precision gains of the linear-rotation protocol
    with a fixed $J_y$-Dicke readout.
    Panels (a) and (b) show the marginal gains $G_{1,\mathrm c}$ and
    $G_{2,\mathrm c}$ for $\theta_1$ and $\theta_2$, respectively.
    In panel (c), the background color represents the total gain
    $G_{\mathrm{tot},\mathrm c}$. The black contours separate regions
    where both marginal gains are positive from those where at least
    one marginal gain is nonpositive.
    All gains are given in decibels relative to the resource-matched
    separated-OAT QFI reference. Positive values indicate classical
    precision bounds below the corresponding reference bounds.
    Here $N=50$, and the preparation and control settings are taken
    from the maximum joint-gain point in Fig.~\ref{fig:CQ_heatmap_rotation}(b).
    These settings and the readout axis remain fixed throughout the maps.
    }
    \label{fig:CFI_rotation}
\end{figure*}
\begin{figure*}
    \centering
    \includegraphics[width=1.0\linewidth]{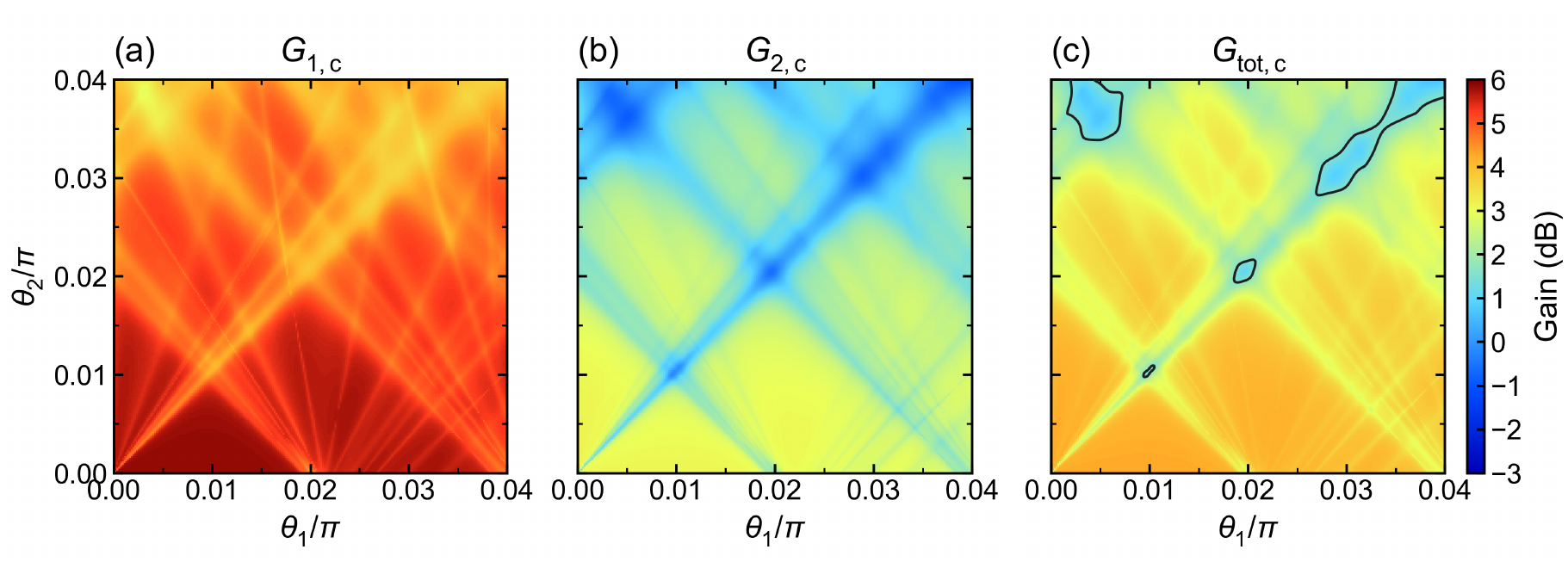}
    \caption{
    Classical precision gains of the intermediate-OAT protocol
    with a fixed $J_x$-Dicke readout.
    Panels (a) and (b) show the marginal gains $G_{1,\mathrm c}$ and
    $G_{2,\mathrm c}$ for $\theta_1$ and $\theta_2$, respectively.
    In panel (c), the background color represents the total gain
    $G_{\mathrm{tot},\mathrm c}$. The black contours delineate the small regions where at least one marginal gain is nonpositive. Both marginal gains are positive over most of the displayed phase region.
    All gains are given in decibels relative to the resource-matched
    separated-OAT QFI reference. Positive values indicate classical
    precision bounds below the corresponding reference bounds.
    Here $N=50$, and the preparation and control settings are taken
    from the maximum joint-gain point in Fig.~\ref{fig:CQ_heatmap}(b).
    Each separated probe is twisted for $t_1+t_2$.
    The preparation, control, and readout settings remain fixed
    throughout the maps.
    }
    \label{fig:CFI}
\end{figure*}

\section{Dicke-basis readout and classical Fisher information}\label{sec:dicke_readout}
The preceding section characterizes the information in the encoded state
and the quantum precision bounds for joint estimation. We now examine
how this information can be extracted with a fixed collective-spin
readout. The CFIM describes the sensitivity of the resulting outcome
probabilities. Even when measurement incompatibility is weak, the
precision of this readout depends on how the two phase responses appear
in those probabilities.

We consider projective measurements of the collective-spin component
$J_{\bm n}=\bm n\cdot\bm J$, where $\bm n$ is a unit vector specifying
the readout axis. The projectors and eigenstates satisfy
\begin{equation}
\begin{aligned}
    \Pi_m^{(\bm n)}
    &=\lvert m\rangle_{\bm n}\,{}_{\bm n}\langle m\rvert,\\[3pt]
    J_{\bm n}\lvert m\rangle_{\bm n}
    &=m\lvert m\rangle_{\bm n},
\end{aligned}
\label{eq:Dicke_projectors}
\end{equation}
with $m=-J,-J+1,\ldots,J$. Such a measurement can be implemented by a
collective rotation followed by a $J_z$-basis measurement. For the
encoded state $\ket{\psi_{\bm\theta}}$, the outcome probabilities are
\begin{equation}
    p_m(\bm\theta)
    =\left|{}_{\bm n}\!\braket{m}{\psi_{\bm\theta}}\right|^2,
    \label{eq:Dicke_probabilities}
\end{equation}
and their phase sensitivity determines the CFIM,
\begin{equation}
    F_{ij}(\bm\theta)
    =\sum_m
    \frac{\partial_i p_m(\bm\theta)\,\partial_j p_m(\bm\theta)}
         {p_m(\bm\theta)}.
    \label{eq:CFIM}
\end{equation}

For nonsingular $F$, the classical precision cost associated with this
measurement is
\begin{equation}
    C_F(W)=\operatorname{Tr}\left(WF^{-1}\right),
    \label{eq:CF_cost}
\end{equation}
which bounds the weighted estimation variance from below and obeys
$C_F(W)\geq C_H(W)\geq C_Q(W)$. Under standard regularity conditions,
this classical bound can be approached asymptotically by repeating the
measurement on identically prepared probes and using an efficient
estimator~\cite{Helstrom_1976,Holevo_1982,Albarelli_2020}.
We take $W=\mathbb I_2$, so that $C_F=\operatorname{Tr}(F^{-1})$ and
$(F^{-1})_{ii}$ gives the marginal bound for $\theta_i$. The full inverse
accounts for the coupling between the two parameters in joint
estimation. As in Sec.~\ref{sec:def}, the common inverse repetition
factor is suppressed.

We compare the fixed-readout precision with the resource-matched
separated-OAT reference introduced in Sec.~\ref{sec:mitigating_slopp}.
This reference consists of two independent $N/2$-particle probes, each
twisted for the total OAT duration of the corresponding joint protocol.
Each branch has single-parameter QFI $q_{\mathrm{sep}}$, giving marginal
and total reference bounds $1/q_{\mathrm{sep}}$ and
$2/q_{\mathrm{sep}}$, respectively. The classical marginal gains
($i=1,2$) and total gain are therefore
\begin{subequations}
\label{eq:CFIM_gain}
\begin{align}
    G_{i,\mathrm c}
    &=-10\log_{10}\left[q_{\mathrm{sep}}(F^{-1})_{ii}\right],
    \label{eq:CFIM_gain_marginal}\\
    G_{\mathrm{tot},\mathrm c}
    &=-10\log_{10}\left[\frac{q_{\mathrm{sep}}C_F}{2}\right].
    \label{eq:CFIM_gain_total}
\end{align}
\end{subequations}
These gains use the separated-OAT reference. A positive marginal gain means
that the fixed joint readout gives a lower classical marginal bound than
the single-parameter QFI bound of the corresponding separated probe,
while positive total gain indicates a lower total bound. Simultaneous
enhancement requires $G_{1,\mathrm c}>0$ and $G_{2,\mathrm c}>0$.
Because the total reference bound is the sum of the two marginal bounds,
these inequalities also ensure positive total gain. A positive total
gain can additionally arise when a larger improvement in one phase
compensates for a deterioration in the other.

Guided by numerical optimization within this class of collective-spin
measurements, we use the $J_y$ eigenbasis for the linear-rotation
protocol and the $J_x$ eigenbasis for the intermediate-OAT protocol.
Figures~\ref{fig:CFI_rotation} and~\ref{fig:CFI} show the fixed-readout
gains for $N=50$ over $0\leq\theta_1,\theta_2\leq0.04\pi$.
The preparation and control settings are taken from the maximum
joint-gain points in Figs.~\ref{fig:CQ_heatmap_rotation}(b)
and~\ref{fig:CQ_heatmap}(b), respectively. Panels (a) and (b) display
$G_{1,\mathrm c}$ and $G_{2,\mathrm c}$, while panel (c) shows
$G_{\mathrm{tot},\mathrm c}$. Each map uses fixed preparation, control,
and readout settings; each point represents local estimation at the
indicated phases. The black contours in panel (c) mark the internal
boundaries of the simultaneous-enhancement region. Although the QFIM is
independent of $\theta_2$, the final rotation changes the measured
populations, so the fixed-readout CFIM generally depends on both phases.

For the linear-rotation protocol, the regions of large $G_{1,\mathrm c}$
and $G_{2,\mathrm c}$ in Figs.~\ref{fig:CFI_rotation}(a)
and~\ref{fig:CFI_rotation}(b) occur at different phase points.
Their positive-gain regions overlap only partially.
Figure~\ref{fig:CFI_rotation}(c) consequently contains regions of
positive total gain where one marginal gain is nonpositive, illustrating
how an improvement in the total cost can coexist with unequal performance
for the two phases. The comparison with the diagonal QFIM of the chosen
linear-control setting highlights the role of the measurement in
determining the balance between the two marginal precisions.

For the intermediate-OAT protocol, Fig.~\ref{fig:CFI}(a) shows positive
gain for $\theta_1$ over a broad phase region. The gain for $\theta_2$
in panel (b) is smaller and includes regions of nonpositive gain, but
the two positive marginal-gain regions overlap over a substantially
broader area than in Fig.~\ref{fig:CFI_rotation}. Panel (c) displays
positive total gain over much of the phase map, with the black contours
marking internal boundaries of simultaneous marginal enhancement.
The fixed $J_x$-Dicke readout thus converts the enhanced quantum
sensitivity of the intermediate-OAT protocol into simultaneous classical
gains over a broad operating region.

The broader simultaneous-enhancement region in Fig.~\ref{fig:CFI} is
useful when the phases are not accurately known in advance. It allows
both marginal gains to remain positive over a wider range of local
working points, easing the requirement for precise preliminary
localization. Finite operating ranges and prior phase uncertainty are
important considerations in dynamic-range and Bayesian quantum
metrology~\cite{Liu_2025_DynamicRange,Kaubruegger_2021}.
In multiparameter estimation, adaptive protocols use preliminary
measurement outcomes to update phase estimates and guide subsequent
controls toward favorable working points~\cite{Ciampini_2016,Valeri_2020}.
The wider operating region provided by intermediate OAT may reduce the
effort needed to locate such a working point, leaving more of the
measurement budget available for precision estimation.

\section{Conclusion and discussion}\label{sec:conclusion}
We have shown that parameter-independent intermediate controls can
restore local distinguishability between two sequential same-axis phase
shifts in a collective-spin interferometer. The essential requirement is
that the two effective generators produce independent local responses on
the prepared probe. A transverse collective rotation separates these
responses by reorienting the second effective generator within the linear
spin operator space, while intermediate OAT introduces nonlinear spin
terms. For the aligned probe and the chosen linear-control setting, the
QFIM is diagonal, with vanishing measurement incompatibility at
$\theta_1=0$ and weak incompatibility nearby.

Under matched particle and twisting resources, both protocols can lower
the marginal SLD bounds for both phases relative to two independent
$N/2$-particle OAT probes. For intermediate OAT, this enhancement extends
over a broad range of control settings in which weak measurement
incompatibility keeps the Holevo bound close to the SLD cost. A separate
analysis based on conditional optimization of the Holevo upper bound
$C_Q(1+\mathcal R)$ numerically demonstrates $N^{-2}$ scaling in the
plateau regime. The characteristic preparation and intermediate-control
scales follow $\chi t_1\sim N^{-1/2}$ and $\chi t_2\sim N^{-1}$,
respectively, showing that this regime can be reached at shorter twisting
durations as the particle number increases.

The fixed-readout analysis connects these quantum precision bounds to
measurements of collective-spin populations. For intermediate OAT, a
fixed $J_x$-Dicke readout lowers the classical marginal bounds for both
phases relative to the separated-OAT QFI reference over a broad region
of the sampled phase domain. The linear protocol with the selected
$J_y$-Dicke readout provides simultaneous enhancement over a smaller
region, despite its diagonal QFIM. This comparison highlights the role
of the measurement in converting the two phase responses into observable
population changes. Favorable quantum sensitivity and a readout that
resolves both responses are therefore complementary requirements for
simultaneous precision enhancement.

Several extensions follow naturally from these results. Nonlinear
readout applied after both encodings may improve robustness to detection
noise~\cite{Davis_2016,Haine_2018,Schulte_2020,Colombo_2022}, with the
measurement designed to retain independent sensitivity to both phases in
its outcome probabilities. For more than two sequential parameters,
repeated intermediate controls could generate additional independent
responses~\cite{Yang2022Untwining,Kaubruegger_2023}, motivating the joint
study of measurement compatibility and readout design. The shared
$\mathrm{SU}(2)$ operator structure also motivates exploring single
large-spin systems with quadratic control~\cite{Chalopin_2018,Yu_2025},
with resource accounting adapted to their internal spin degrees of
freedom. These results show how nonlinear intermediate control can
restore parameter distinguishability and make simultaneous precision
gains accessible through a common fixed readout.

\section*{Acknowledgments}
This work is supported by the
National Key R\&D Program of China (Grant
No. 2025YFA1411600), by the National Natural Science Foundation of China (NSFC)
(Grants No.~12304543, No.~92476205, No.~12361131576 and No.~92265205), and by the Quantum Science and Technology-
National Science and Technology Major Project (Grant No.2021ZD0302104). R. L. is supported by the“Gravitational Wave Detection” program (No. 2023YFC2205800) funded by the Ministry of Science and Technology of the People's Republic of China.

\appendix
\section{Analytic results for the linear-rotation protocol}\label{app:linear_rotation}
We give the linear-rotation results at arbitrary $\theta_1$, writing
$\gamma=\beta-\theta_1$. To make the role of probe alignment explicit,
we initially leave the preparation angle $\delta$ free. The probe's
$\pi$-rotation symmetry about $x$ gives
$\langle J_y\rangle_P=\langle J_z\rangle_P=0$ and vanishing real
covariances between $J_x$ and the other two spin components. Retaining
the $yz$ covariance, the QFIM, the independent Uhlmann-curvature
component, and the SLD cost are
\begin{widetext}
\begin{equation}
    Q = 4
    \begin{pmatrix}
        \mathcal V_z & \mathcal V_z\cos\alpha+\mathcal C_{zy}\sin\alpha\cos\gamma \\
        \mathcal V_z\cos\alpha+\mathcal C_{zy}\sin\alpha\cos\gamma &
        \mathcal V_z\cos^2\alpha
        +\sin^2\alpha(\mathcal V_x\sin^2\gamma+\mathcal V_y\cos^2\gamma)
        +2\mathcal C_{zy}\sin\alpha\cos\alpha\cos\gamma
    \end{pmatrix},
    \label{eq:Q_linear_rotation}
\end{equation}
\begin{equation}
    U_{12}=-2\langle J_x\rangle_P\sin\alpha\cos\gamma,
    \label{eq:U12_linear_rotation}
\end{equation}
\begin{equation}
    C_Q = \frac{\mathcal V_z(1+\cos^2\alpha)
    +\sin^2\alpha(\mathcal V_x\sin^2\gamma+\mathcal V_y\cos^2\gamma)
    +2\mathcal C_{zy}\sin\alpha\cos\alpha\cos\gamma}
    {4\sin^2\alpha[\mathcal V_z\mathcal V_x\sin^2\gamma
    +(\mathcal V_z\mathcal V_y-\mathcal C_{zy}^2)\cos^2\gamma]}.
    \label{eq:CQ_linear_rotation}
\end{equation}
\end{widetext}
All expectation values and covariances are evaluated in
$\ket{\psi_P}$, with
\begin{equation}
    \mathcal V_i=\operatorname{Var}_P(J_i),
    \qquad i=x,y,z,
    \label{eq:linear_variances_app}
\end{equation}
\begin{equation}
    \mathcal C_{zy}=\operatorname{Re}\operatorname{Cov}_P(J_z,J_y).
    \label{eq:linear_covariance_app}
\end{equation}
The cost in Eq.~\eqref{eq:CQ_linear_rotation} is defined where the
QFIM is nonsingular.

At fixed $\gamma$, Eq.~\eqref{eq:CQ_linear_rotation} is quadratic in
$\cot\alpha$. Minimizing this quadratic and then varying $\gamma$
gives the representative minimizers
\begin{equation}
\begin{aligned}
    \cot\alpha&=-\frac{\mathcal C_{zy}\cos\gamma}{2\mathcal V_z},\\[2pt]
    \gamma&=\begin{cases}
        \pi/2, & \Xi>0,\\
        0,\pi, & \Xi<0,
    \end{cases}
\end{aligned}
    \label{eq:linear_minimizers_app}
\end{equation}
where
\begin{equation}
    \Xi=\mathcal V_z^2(\mathcal V_x-\mathcal V_y)
    +\mathcal C_{zy}^2\left(\mathcal V_z+\frac{\mathcal V_x}{2}\right).
    \label{eq:linear_minimum_indicator_app}
\end{equation}
If $\Xi=0$, the cost minimized over $\alpha$ is independent of
$\gamma$ wherever the QFIM remains nonsingular.

The alignment rotation used in the main text diagonalizes the real
covariance matrix in the $J_y$--$J_z$ plane, so that
\begin{equation}
    \mathcal C_{zy}=0.
    \label{eq:aligned_covariance_app}
\end{equation}
Equation~\eqref{eq:CQ_linear_rotation} then reduces to
\begin{equation}
    C_Q=\frac{1+\cos^2\alpha}
    {4\sin^2\alpha(\mathcal V_x\sin^2\gamma+\mathcal V_y\cos^2\gamma)}
    +\frac{1}{4\mathcal V_z}.
    \label{eq:aligned_linear_cost_app}
\end{equation}
The minimizing angles can be chosen as
\begin{equation}
\begin{aligned}
    \alpha&=\frac{\pi}{2},\\[2pt]
    \gamma&=\begin{cases}
        \pi/2, & \mathcal V_x>\mathcal V_y,\\
        0,\pi, & \mathcal V_x<\mathcal V_y.
    \end{cases}
\end{aligned}
    \label{eq:aligned_linear_minimizers_app}
\end{equation}
When $\mathcal V_x=\mathcal V_y$, all $\gamma$ giving a nonsingular
QFIM are equivalent. Integer multiples of $\pi$ can be added to
$\alpha$ and $\gamma$ without changing this aligned-probe cost.

In the preparation regime $\mathcal V_x>\mathcal V_y$, the fixed
setting $\alpha=\beta=\pi/2$ is locally optimal at $\theta_1=0$.
For this setting, the QFIM at arbitrary $\theta_1$ is
\begin{equation}
    Q=4
    \begin{pmatrix}
        \mathcal V_z & 0\\
        0 & \mathcal V_x\cos^2\theta_1+\mathcal V_y\sin^2\theta_1
    \end{pmatrix},
    \label{eq:Q_linear_rotation_diagonal_app}
\end{equation}
and the independent curvature component is
\begin{equation}
    U_{12}=-2\langle J_x\rangle_P\sin\theta_1.
    \label{eq:aligned_linear_curvature_app}
\end{equation}
Thus, the QFIM remains diagonal as $\theta_1$ varies, while the
measurement incompatibility vanishes at $\theta_1=0$ and is weak nearby.

\section{Twisting-axis dependence of the intermediate-OAT protocol}\label{app:twisting}
\begin{figure}
    \centering
    \includegraphics[width=1.0\linewidth]{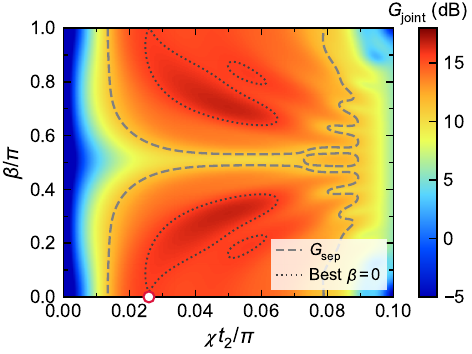}
    \caption{
    Twisting-axis dependence of the intermediate-OAT protocol.
    The color map shows the joint SLD precision gain $G_{\mathrm{joint}}$,
    relative to the separated SQL, versus the intermediate twisting
    strength $\chi t_2/\pi$ and axis angle $\beta/\pi$.
    The twisting axis lies in the $xy$ plane, with $\beta$ measured
    from the $x$ axis. The gray dashed contour labeled
    $G_{\mathrm{sep}}$ marks equality with the resource-matched
    separated-OAT reference,
    $G_{\mathrm{joint}}=G_{\mathrm{sep}}=11.26\,\mathrm{dB}$.
    The gray dotted contours labeled ``Best $\beta=0$'' mark
    $G_{\mathrm{joint}}=15.56\,\mathrm{dB}$, the maximum sampled gain
    along the $\beta=0$ ($x$-axis) cut. Gains above these two levels
    correspond to lower total SLD costs than the separated-OAT reference
    and the best sampled $x$-axis protocol, respectively.
    The magenta open circle identifies the best sampled point on the
    $\beta=0$ cut, corresponding to the representative $x$-axis working
    point used in Fig.~\ref{fig:CQ_heatmap}(b).
    }
    \label{fig:twisting_axis_scan}
\end{figure}
A general intermediate OAT control about an axis in the $xy$ plane is
\begin{equation}
    V=T_\beta(t_2)=e^{-i\chi t_2J_\beta^2},
    \label{eq:V2_twisting_general}
\end{equation}
where
\begin{equation}
    J_\beta=J_x\cos\beta+J_y\sin\beta,
    \label{eq:twisting_axis_app}
\end{equation}
and $\beta$ is measured from the $x$ axis. Conjugation by
$e^{i\theta_1J_z}$ maps $J_\beta$ to $J_{\beta-\theta_1}$, so the
second local generator can be written exactly as
\begin{align}
    \mathcal H_2
    &=e^{i\theta_1J_z}e^{i\chi t_2J_\beta^2}J_z
      e^{-i\chi t_2J_\beta^2}e^{-i\theta_1J_z}
      \label{eq:twisting_local_generator_app}\\
    &=e^{i\chi t_2J_\gamma^2}J_z e^{-i\chi t_2J_\gamma^2},
      \label{eq:twisting_rotated_generator_app}
\end{align}
with $\gamma=\beta-\theta_1$, as in
Appendix~\ref{app:linear_rotation}. Near $\theta_1=0$,
$\gamma\simeq\beta$, so the controlled angle $\beta$ closely tracks
the effective twisting direction. The numerical scan uses the exact local
generator.

Figure~\ref{fig:twisting_axis_scan} shows the joint gain as both the
intermediate twisting strength and axis are varied. The best sampled
point lies near $\beta\simeq0.3\pi$, while high gain extends over a
broad range of axis directions. Along the representative $\beta=0$
cut, the maximum sampled gain is $15.56\,\mathrm{dB}$, above the
resource-matched separated-OAT reference of $11.26\,\mathrm{dB}$.
We therefore use the $x$-axis control in the main text: it provides
enhanced joint precision and has the simple leading weak-twisting
correction $\chi t_2\{J_x,J_y\}$ at $\theta_1=0$, as derived in
Appendix~\ref{app:weak_oat}.

\section{Weak-twisting expansion and QFIM rank restoration}\label{app:weak_oat}
We derive the weak-twisting mechanism behind the rank lifting in
Sec.~\ref{subsec:nonlinear}. For the intermediate control
$V=e^{-i\chi t_2J_x^2}$, the local generators are
$\mathcal H_1=J_z$ and
\begin{equation}
    \mathcal H_2=e^{i\theta_1J_z}e^{i\chi t_2J_x^2}J_z
    e^{-i\chi t_2J_x^2}e^{-i\theta_1J_z}.
    \label{eq:weak_twisting_generator_app}
\end{equation}
Setting $\tau=\chi t_2$ and evaluating at $\theta_1=0$, the
Baker--Campbell--Hausdorff expansion for $|\tau|J\ll1$ gives
\begin{equation}
\begin{aligned}
    \mathcal H_2
    &=J_z+\tau\{J_x,J_y\}\\
    &\quad-\frac{\tau^2}{2}\{J_x,\{J_x,J_z\}\}+O(\tau^3).
\end{aligned}
    \label{eq:weak_twisting_expansion_app}
\end{equation}
The first correction introduces the quadratic operator
$K_1=\{J_x,J_y\}$, while the second involves the cubic operator
$K_2=\{J_x,\{J_x,J_z\}\}$. Both corrections vanish when $t_2=0$.

All moments are evaluated in the prepared probe $\ket{\psi_P}$.
Defining
\begin{align}
    \mathcal V_z&=\operatorname{Var}_P(J_z),
    \label{eq:weak_variance_z_app}\\
    \mathcal V_{K_1}&=\operatorname{Var}_P(K_1),
    \label{eq:weak_variance_k1_app}\\
    \mathcal C_{zK_1}&=\operatorname{Re}\operatorname{Cov}_P(J_z,K_1),
    \label{eq:weak_covariance_k1_app}\\
    \mathcal C_{zK_2}&=\operatorname{Re}\operatorname{Cov}_P(J_z,K_2),
    \label{eq:weak_covariance_k2_app}
\end{align}
we expand the real covariances to second order in $\tau$ to obtain
\begin{align}
    Q_{11}/4&=\mathcal V_z,
    \label{eq:weak_q11_app}\\
    Q_{12}/4&=\mathcal V_z+\tau\mathcal C_{zK_1}\notag\\
    &\quad-\frac{\tau^2}{2}\mathcal C_{zK_2}+O(\tau^3),
    \label{eq:weak_q12_app}\\
    Q_{22}/4&=\mathcal V_z+2\tau\mathcal C_{zK_1}\notag\\
    &\quad+\tau^2(\mathcal V_{K_1}-\mathcal C_{zK_2})+O(\tau^3).
    \label{eq:weak_q22_app}
\end{align}
At $t_2=0$, the two local generators coincide and the QFIM has rank
one. The terms involving $K_2$ cancel from the determinant at order
$\tau^2$, leaving
\begin{equation}
    \det Q=16\tau^2
    \left(\mathcal V_z\mathcal V_{K_1}-\mathcal C_{zK_1}^2\right)
    +O(\tau^3).
    \label{eq:weak_determinant_app}
\end{equation}
The expression in parentheses is the Gram determinant of the centered
observables $J_z$ and $K_1$ in the real covariance metric. It is
positive when their fluctuations on the probe are linearly independent,
so a sufficiently weak nonzero intermediate twist restores full rank.
Intermediate OAT therefore separates the two phase responses by
introducing a nonlinear generator direction. The numerical results in
the main text are evaluated using the exact local generators.

\end{document}